4# Description of electromagnetic fields in inhomogeneous accelerating sections. I Field representation

4444
M.I. Ayzatsky

National Science Center Kharkov Institute of Physics and Technology (NSC KIPT), 61108, Kharkov, Ukraine

E-mail: mykola.aizatsky@gmail.com

In this work we present the results of a study of the possibility of using a homogeneous basis and a new generalization of coupled modes theory to describe inhomogeneous accelerating sections. It was shown that the single mode representation describes the electromagnetic fields with small errors in the regular part of the section. In the couplers the accuracy is worse, but it is small enough to be used in calculations. Solving the equations of generalized coupled modes theory requires calculating the derivatives of the fields with respect to various geometrical parameters. Procedures for calculating these derivatives have been developed.

## 1. INTRODUCTION

Recently it was proposed to use a uniform basis for description of non-periodic structured waveguides and a new generalization of the theory of coupled modes was constructed [1]. The obtained infinitive system of coupled equations transfers into the known system for periodic waveguides as the inhomogeneity tends to zero [2]. Before using proposed system of coupled equations to calculate waveguide characteristics, it is necessary to solve two problems: a) establish the possibility of reducing an infinitive system of equations and b) develop the procedures for numerical calculation the coefficients that are needed for solving the coupling equations.

The best way to solve the first problem (a) is to calculate the waveguide field distribution based on a rigorous electromagnetic code, find the expansion coefficients, and estimate how many terms need to be included to match the two field distributions and what is the accuracy of the matching.

This procedure requires solving two tasks: calculating the field distribution in non-periodic structured waveguides and calculating the eigen characteristics (propagation constants and eigen vectors) of periodic waveguides for various geometrical parameters.

The results of the preliminary calculations showed that the single wave approximation gives good results for slow changes in the waveguide dimensions [3]. But the accelerator sections include the couplers and the section dimensions change relatively fast in these segments. This fact demands more deep investigation of possibility of using the coupling equations.

For solution of the second problem (b) it is necessary to develop procedures for calculating the fields derivatives with respect to different geometrical parameters.

For the case of Disc-Loaded Waveguide (DLW) the Coupled Integral Equations Method (CIEM) (see, for example, [4,5,6,7]) is the most suitable, since the mentioned above problems can be solved on its bases [8]. The code CASCIE (Code for Accelerating Structures - Coupled Integral Equations) was used for calculation the field distribution in non-periodic structures [9]. We also developed a code that calculate the eigen characteristics of periodic DLWs. We used IMSL Fortran Numeric Library for computing the eigenvalues and eigenvectors of a complex matrices. We also developed procedures for calculating the fields derivatives with respect to different geometrical parameters.

## 2. MODEL OF ACCELERATING SECTION

The accelerator section being developed at CERN [10] was taken as a prototype. The main characteristics of a model section are presented in Table 1. The structure of the model section and main designations are shown on Figure 1 and Figure 2.

The main difference between the model section and the CERN section is the type of couplers. In CERN section the magnetic couplers are used, in the model section – the electric couplers [11,12].

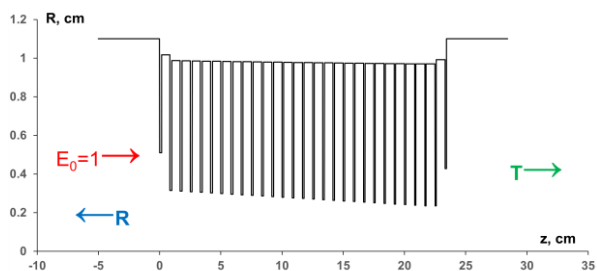

*Figure 1 The structure of the model section*

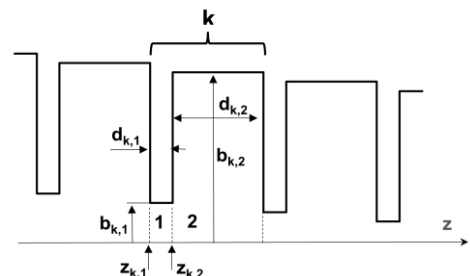

*Figure 2 Designations that are used in the article*

Table 1 Structure Parameters

|  |  | CERN [10] |
|---|---|---|
| Frequency, GHz | 11.994 | 11.994 |
| RF phase advance per cell, rad. | $2\pi/3$ | $2\pi/3$ |
| Input, Output iris radii ($b_{k,1}$), mm | 3.15, 2.35 | 3.15, 2.35 |
| Input, Output iris thickness ($d_{k,1}$), mm | 1.67, 1.00 | 1.67, 1.00 |
| Structure period $D = d_{k,1} + d_{k,2}$, mm | 8.3317 | 8.3317 |
| Input, Output group velocity, % of c | 1.44, 0.76 | 1.44, 0.76 |
| First and last cell Q-factor | 5560, 5560 | 5536, 5738 |
| Number of regular cells | 26 | 26 |

### 3. FIELD EXPANSION

We will consider only axially symmetric TM fields with $E_z, E_r, H_\varphi$ components. Time dependence is $\exp(-i\omega t)$.

Electromagnetic fields in a non-periodic structured waveguide with the ideal metal walls can be represented in the form of such series (see Appendix 1 and [1])

$$\vec{H}(\vec{r}) = \sum_{s=-\infty}^{s=\infty} C_s(z) \vec{H}_s^{(e,z)}(\vec{r}), \qquad (1)$$

$$\vec{E}(\vec{r}) = \sum_{s=-\infty}^{s=\infty} C_s(z) \vec{E}_s^{(e,z)}(\vec{r}) + \frac{\vec{j}_z}{i\omega\varepsilon_0\varepsilon}, \qquad (2)$$

where $\vec{E}_s^{(e,z)}(\vec{r}), \vec{H}_s^{(e,z)}(\vec{r})$ are modified eigen vector functions obtained by generalizing the eigen $\vec{E}_s^{(e)}, \vec{H}_s^{(e)}$ vectors of a homogeneous waveguide by special continuation of the geometric parameters (see Appendix 2 and [1,3]). The eigen waves of a homogeneous waveguide we present as $(\vec{E}_s, \vec{H}_s) = (\vec{E}_s^{(e)}, \vec{H}_s^{(e)}) \exp(\gamma_s z)$, where ($\vec{E}_s^{(e)}, \vec{H}_s^{(e)}$) are the periodic functions of the z-coordinate. Under such choice of the basis functions, the coefficients $C_s(z)$ include an exponential dependence on the z-coordinate and, strictly speaking, cannot be named as amplitude. But we will use this term for the sake of simplicity of the further presentation.

Since $\vec{E}_s^{(e,z)}(\vec{r}), \vec{H}_s^{(e,z)}(\vec{r})$ are orthogonal [1], from (1) and (2) we obtain

$$N_s^{(e,z)} C_s(z) = \int_{S_\perp^{(z)}(z)} \left( \left[ \left( \vec{E} - \frac{\vec{j}_z}{i\omega\varepsilon_0\varepsilon} \right) \vec{H}_{-s}^{(e,z)} \right] - \left[ \vec{E}_{-s}^{(e,z)} \vec{H} \right] \right) \vec{e}_z dS, \qquad (3)$$

where

$$N_s^{(e,z)}(z) = \int_{S_t} \left( \left[ \vec{E}_s^{(e,z)} \vec{H}_{-s}^{(e,z)} \right] - \left[ \vec{E}_{-s}^{(e,z)} \vec{H}_s^{(e,z)} \right] \right) \vec{e}_z dS \qquad (4)$$

Electromagnetic fields (1)-(2) must obey the Maxwell's equations. It can be shown that $C_s(z)$ must be solutions of such a coupled system of differential equations (see Appendix 1 and [1])

$$\frac{dC_s}{dz} - \gamma_s^{(e,z)} C_s + \frac{1}{2N_s^{(e,z)}} \frac{dN_s^{(e,z)}}{dz} C_s + \sum_{s'=-\infty}^{\infty} C_{s'} U_{s',s}^{(z)} = \frac{1}{N_s^{(e,z)}} \int_{S_\perp^{(z)}} \vec{j} \vec{E}_{-s}^{(e,z)} dS, \qquad (5)$$

where

$$U_{k',k}^{(z)} = \frac{1}{2N_s^{(e,z)}} \sum_i \frac{dg_i^{(z)}}{dz} \int_{S_\perp^{(z)}(z)} \left\{ \left[ \frac{\partial \vec{E}_{k'}^{(e,z)}}{\partial g_i^{(z)}} \vec{H}_k^{(e,z)} \right] + \left[ \frac{\partial \vec{E}_k^{(e,z)}}{\partial g_i^{(z)}} \vec{H}_{k'}^{(e,z)} \right] - \left[ \vec{E}_k^{(e,z)} \frac{\partial \vec{H}_{k'}^{(e,z)}}{\partial g_i^{(z)}} \right] - \left[ \vec{E}_{k'}^{(e,z)} \frac{\partial \vec{H}_k^{(e,z)}}{\partial g_i^{(z)}} \right] \right\} \vec{e}_z dS, \qquad (6)$$

$g_i^{(z)}(z)$ -generalized geometrical parameters, $\gamma_s^{(e,z)}(z)$ - generalized wavenumber (see Appendix 2 and [1,3]).

In this paper we will consider axisymmetric TH (E) electromagnetic fields in the model section without current ($\vec{j} = 0$). We will introduce the notion of the elementary cell with a number $k$ (see Figure 2), which starts at $z_{k,1} = \sum_i^{k-1}(d_{i,1} + d_{i,2})$ and consists of the first segment of a circular waveguide (region 1, $z_{k,1} < z < z_{k,1} + d_{k,1} = z_{k,2}$, length $d_{k,1}$, radius of waveguide $b_{k,1}$) and the second segment of a circular waveguide (region 2, $z_{k,2} < z < z_{k+1,1}$, radius $b_{k,2}$, length $d_{k,2}$).

3In CIEM electric and magnetic fields in the region $q$ ($q=1,2$) are represented in the form of series [4,5,8]

$$E_r^{(k,q)} = \sum_m \left\{ B_{m,1}^{(k,q)} \exp\left(\chi_m^{(k,q)} \tilde{z}\right) + B_{m,2}^{(k,q)} \exp\left(-\chi_m^{(k,q)} \tilde{z}\right) \right\} J_1\left(\frac{\lambda_m}{b_{k,q}} r\right), \quad (7)$$

$$E_z^{(k,q)} = -\sum_m \frac{\lambda_m}{\chi_m^{(k,q)} b_{k,q}} \left\{ B_{m,1}^{(k,q)} \exp\left(\chi_m^{(k,q)} \tilde{z}\right) - B_{m,2}^{(k,q)} \exp\left(-\chi_m^{(k,q)} \tilde{z}\right) \right\} J_0\left(\frac{\lambda_m}{b_{k,q}} r\right), \quad (8)$$

$$H_\varphi^{(k,q)} = i \frac{\omega b_{k,q}}{c} \frac{1}{Z_0} \sum_m \frac{\varepsilon}{\chi_m^{(k,q)} b_{k,q}} \left\{ B_{m,1}^{(k,q)} \exp\left(\chi_m^{(k,q)} \tilde{z}\right) - B_{m,2}^{(k,q)} \exp\left(-\chi_m^{(k,q)} \tilde{z}\right) \right\} J_1\left(\frac{\lambda_m}{b_{k,q}} r\right), \quad (9)$$

where $\left(\chi_m^{(k,q)}\right)^2 = \left(\frac{\lambda_m}{b_{k,q}}\right)^2 - \left(\frac{\omega}{c}\right)^2 \varepsilon$, $\tilde{z} = z - z_{k,q}$.

Coefficients $B_{m,1}^{(k,q)}, B_{m,2}^{(k,q)}$ are constant inside each segment and are found by equating the fields on the dividing surfaces. In addition to the standard division of the structured waveguide by interfaces between the adjacent regions, we use additional interfaces in places where electric field has the simplest transverse structure [8]. Moreover, the system of coupled integral equations is formulated for longitudinal electrical fields in contrast to the standard approach where the transverse electrical fields are unknowns. The final equations are a system of coupling matrix equations for expansion coefficients of the longitudinal electric field at these additional interfaces. Knowing expansion coefficients, we can calculate $B_{m,1}^{(k,q)}, B_{m,2}^{(k,q)}$ and, therefore, the field distribution.

Such approach gives us a simple procedure for finding eigen vectors of homogeneous waveguides. Instead of solving the system of coupling matrix equations, we must solve the eigen matrix problem, find eigen values and matrix eigen vectors and for selected eigen vector find coefficients $B_{m,1}^{(q,s)}, B_{m,2}^{(q,s)}$. For solving the eigen matrix problem we used IMSL Fortran Numeric Library Routine EVCCG.

The field distribution for the eigen wave we find using expressions (7)-(9) where $B_{m,1}^{(k,q)}, B_{m,2}^{(k,q)}$ are changed to $B_{m,1}^{(q,s)}, B_{m,2}^{(q,s)}$. After introducing a continuation of the geometric parameters, we can find $C_s(z), C_{-s}(z)$ (see [1,3]). Since $J_1(\lambda_m r / b_{k,q})$ is the orthogonal set of functions, the integrals in (3) are converted into sums.

A similar method was used for calculating coupling coefficients $U_{1,1}, U_{-1,-1}, U_{-1,1}, U_{1,-1}$. Derivatives of $B_{m,1}^{(q,s)}, B_{m,2}^{(q,s)}$ with respect to geometrical parameters were calculated numerically.

Using the same approach to calculate the distributions of fields and eigen waves make it possible to reduce the influence of calculation errors on the results of assessing the accuracy of the decomposition (1)-(2)

As there are difficulties in solving eigen matrix problem with large (small) eigen values (evanescent fields), we will be interesting in possibility of using only two terms in the sums (1) and (2) – $s=1,-1$, when we have a single-mode representation

$$\vec{E}_1(\vec{r}) = \vec{E}_1^+(\vec{r}) + \vec{E}_1^-(\vec{r}) = C_1(z)\vec{E}_1^{(e,z)}(\vec{r}) + C_{-1}(z)\vec{E}_{-1}^{(e,z)}(\vec{r}). \quad (10)$$

Comparing results of calculation based on this approximation with the results obtained on the basis of coupling matrix equations [9], we can make the conclusion of possibility of using the single mode description (10).

We investigated the electromagnetic field expansion (1),(2) for model structure which consist of 26 regular cells and two coupler cells: one upstream (input) coupler and the other downstream (output) coupler. The couplers are connected to input and output infinitive cylindrical waveguides [9]. We also suppose that in the left semi-infinite waveguide the $TM_{0,1}$ eigen wave with unit amplitude $E_z = J_0(\lambda_{01} r / b_w) \exp(i h_w z)$ propagates towards the considered section. Bellow all field values will correspond to this amplitude.

Energy absorption in the walls was modeled by introducing a complex dielectric into each cell ($\varepsilon =$ 1+i1.8E-4) [8,9]. This value of complex permeability gives the required value of quality factor $Q \sim 5600$.

The input and output couplers were tuned to homogeneous DLWs with irises radii and thicknesses $b_{2,1} = ... = b_{28,1} = 3.15$ mm , $d_{1,1} = ... d_{29,1} = 1.67$ mm (type I) and $b_{2,1} = ... = b_{28,1} = 2.35$ mm , $d_{1,1} = ... d_{29,1} = 1$ mm (type II), respectively. The tuning procedure of the output coupler was based on the property of fields in a homogeneous DLW, which consists in the fact that the phase shift between cells is constant. Therefore, if we can find $b_{29,1}$ and $b_{28,2}$ when $\Delta\varphi_k = \Delta\varphi_{k+1} = 2\pi/3$, we can assume that the amplitude of the reflected from coupler wave equals zero. Usually, we stopped iterations if $|\Delta\varphi_{k+1} - \Delta\varphi_k| < 1E-2$ degree.

It is known that in lossy DLWs the input coupler has slightly different geometric dimensions than the output coupler [13,14]. Therefore, only after tuning the output coupler the input coupler can be tuned by minimizing the reflectance of the entire structure. Two sets of couplers were tuned and such reflection coefficients ($|S_{1,1}|$) from the input couplers were achieved: $R_{in} \approx 1.1E-3$ (I) and $R_{in} \approx 8E-4$ (II).

3.1 Homogeneous section

The term "homogeneous accelerating section" usually used to refer to a section in which the regular cells are identical. But due to the existence couplers they are indeed nonuniform. There are only a few publications that attempt to describe coupler fields semi-analytically. Only numerical methods are used to calculate the field distributions in the couplers [11,12,15].

Results of couplers tuning and accuracy of the single mode description (10) for the "homogeneous accelerating section" of type II are presented in Figure 3-Figure 15. In all the figures $D = d_{k,1} + d_{k,2} = const$.

Spatial distributions of the longitudinal electric field $E_z(r=0,z)$, calculated using the CASCIE code, are presented in Figure 3 and Figure 4.

The main role in the studied representation is played by modified eigenmodes $\vec{E}_{\pm 1}^{(e,z)}(\vec{r})$. They were calculating with such normalization condition: $\left|E_{z,\pm 1}^{(e,z)}(r=0, z=d_{c,1}+d_{c,2}/2)\right|=1$, where $d_{c,1}$ is the diaphragm thickness, $d_{c,2}$ - the resonator length of the considered eigen mode (see Figure 2). For a homogeneous waveguides, the distribution of $\vec{E}_{\pm 1}^{(e,z)}(r=0,z)$ is periodic in the region of regular cells, whereas in the region of couplers its distribution is very different (see Figure 5). The reason for this will become clear below.

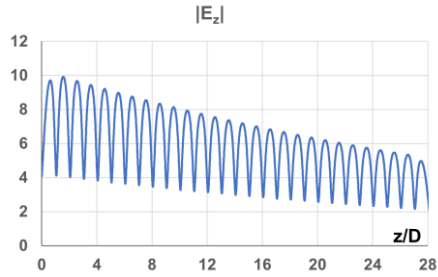

Figure 3 Spatial distributions of the modulus of the longitudinal electric field $E_z(r=0,z)$ (CASCIE code)

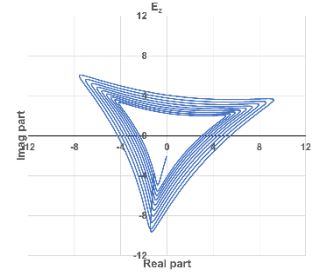

Figure 4 Complex longitudinal electric field $E_z(r=0,z)$ along the axis z (CASCIE code)

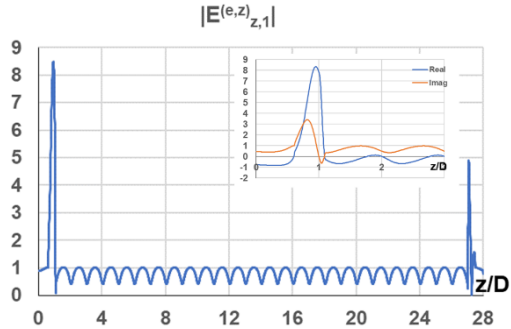

Figure 5 Distribution of longitudinal electric field of modified eigen mode (1-real part, 2- imag part)

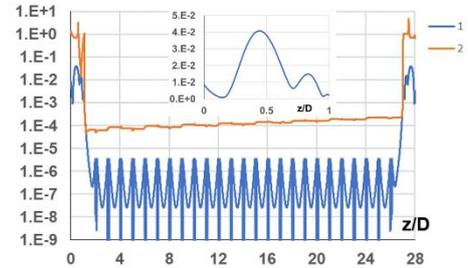

Figure 6 Modulus of relative error in the representation of the longitudinal electric field $E_z(r=0,z)$, calculated using the CASCIE code, by the field of the single mode approach $E_{z,1}^+(r=0,z)$ (1) and by the field $E_{z,1}^+(r=0,z)$ (2)

In the region of regular cells, the relative error in representing the longitudinal electric field $E_z(r=0,z)$, calculated using the CASCIE code, by the single-mode approximation field is less than 1.E-5 (see Figure 5). The relative error in representing the field $E_z(r=0,z)$ by the field $E_{z,1}^+(r=0,z)$ is determined by the reflection from the output coupler (see Figure 6).

In the coupler cells the single mode description gives the error up to 4%. To increase this accuracy, it is necessary to take into account the higher modes. The accuracy of the representation of $E_z(r=0,z)$ by only the field $E_{z,1}^+(r=0,z)$ is very low. This circumstance shows that a correct description of the field distribution over the entire "homogeneous" section is impossible without taking into account the $E_{z,1}^-$ component.

Results of calculations using the CASCIE code show that the distribution of electric field $E_z(r=0,z)$ in the coupler region is smooth and slow (see Figure 7-a). This field we are representing as the sum of two orthogonal components $E_{z,1}^{+}(r=0,z)$ and $E_{z,1}^{-}(r=0,z)$: $E_z = E_{z,1}^{+} + E_{z,1}^{-}$. Dependences on $z$ of these components (Figure 7 b, c). strongly differs from the dependence of their sum (Figure 7 a). Such dependences due to the fact that in some part of the section the modified eigen vectors $E_{z,1}^{+}$ and $E_{z,1}^{-}$ become "evanescent" – their wave numbers become reals (see Figure 8 and Figure 9). There are large differences between the size of the holes in the first and second diaphragms, and between the radii of the first and second resonators (see Figure 1). Inside the coupler resonator we are making smooth transition between these values. At the beginning of resonator modified eigen modes have a phase shift $\mathrm{Im}\left(D\gamma_s^{(e,z)}\right)$ of about 1 rad., then as the effective aperture decreases, the phase shift increases and at some point $\mathrm{Im}\left(D\gamma_s^{(e,z)}\right)$ becomes equal $\pi$ pad. Then the wave number becomes real and negative for $E_{z,1}^{+}$ and positive for $E_{z,1}^{-}$. The effective aperture size stops changing at the end of the resonator, and inside the second diaphragm the resonator radius changes. After the second diaphragm the eigen vectors $E_{z,1}^{+}$ and $E_{z,1}^{-}$ become "propagating" with a phase shift $\mathrm{Im}\left(D\gamma_s^{(e,z)}\right) = 2\pi/3$.

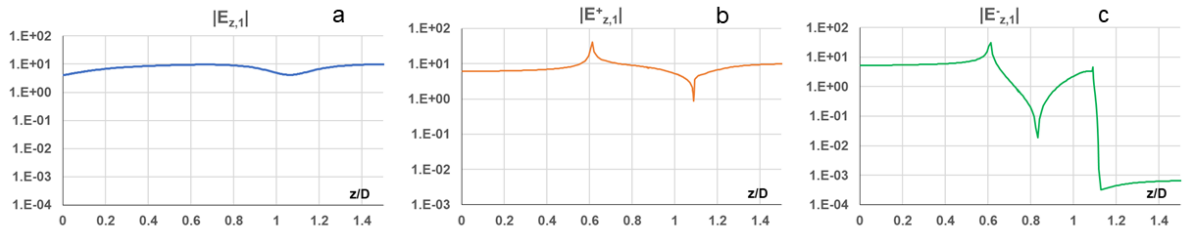

*Figure 7 Dependences on $z$ of various components of the longitudinal electric field*

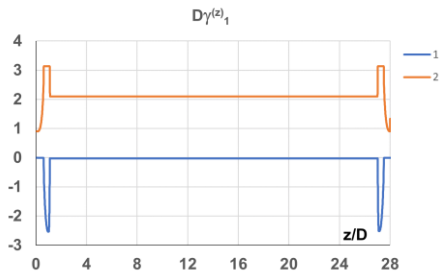

*Figure 8 Dependence of phase shift $D\gamma_s^{(e,z)}$ on coordinate $z$ along the entire section; 1-real part, 2-imaginary part*

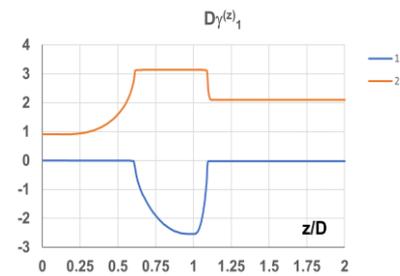

*Figure 9 Dependence of phase shift $D\gamma_s^{(e,z)}$ on coordinate $z$ along the input coupler; 1-real part, 2-imaginary part*

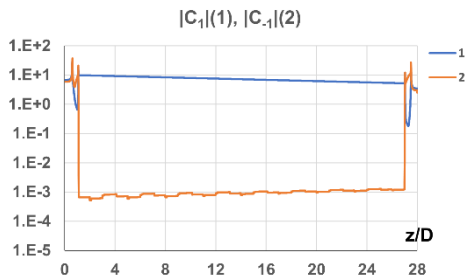

*Figure 10 Spatial distributions of the modulus of amplitudes $C_1$ (1) and $C_{-1}$ (2)*

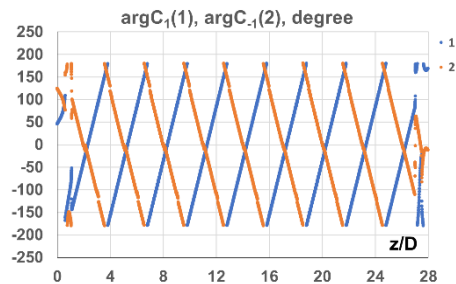

*Figure 11 Spatial distributions of the argument of amplitudes $C_1$ (1) and $C_{-1}$ (2)*

Distribution of the modules and phases of $C_1(z)$ and $C_{-1}(z)$ in the homogeneous part of the section coincides with the well-known physical model of wave propagation (see Figure 10 and Figure 11). There are forward and backward waves, decreasing from the point of their origin. For the forward (backward) wave phase linearly increase (decrease) along the section with phase shift per cell $2\pi/3$ ($-2\pi/3$)

Behavior of the amplitudes $C_1(z)$ and $C_{-1}(z)$ inside the coupler cells differs from the behavior in the regular part (see Figure 12 and Figure 13). This resembles a pattern of standing waves.

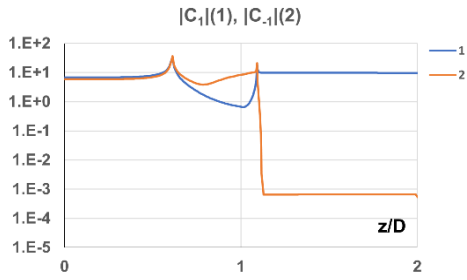
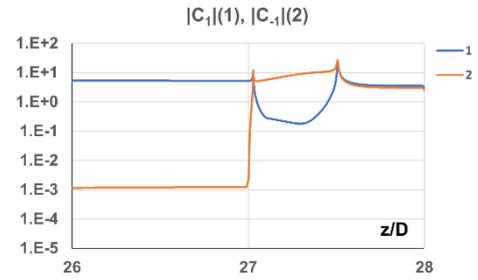

Figure 12 Spatial distributions of the modulus of amplitudes $C_1$ (1) and $C_{-1}$ (2) in the input coupler region

Figure 13 Spatial distributions of the modulus of amplitudes $C_1$ (1) and $C_{-1}$ (2) in the output coupler region

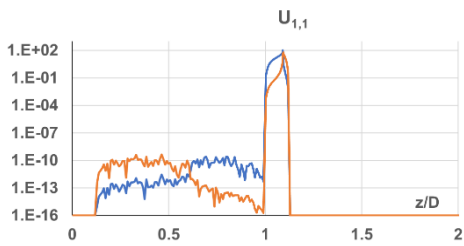
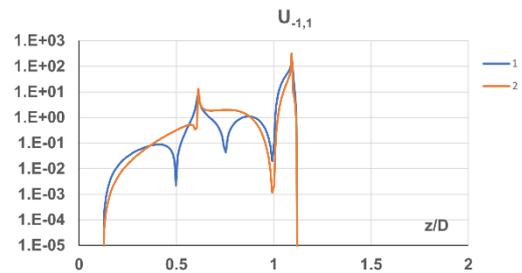

Figure 14 Dependence of $U_{1,1}$ on $z$ in the input coupler; 1-real part, 2- imaginary part

Figure 15 Dependence of $U_{-1,1}$ on $z$ in the input coupler; 1-real part, 2- imaginary part

To use the coupled mode theory(5) to describe a "homogeneous" section, we need to know the coupling coefficients $U_{s',s}$. As follows from (6) on the homogeneous part $dg_i^{(z)}/dz = 0$ and $U_{s',s} = 0$. Dependencies of $U_{s',s}$ on $z$ in the input coupler are given in Figure 14 and Figure 15. $U_{1,1}$ ( $U_{-1,-1} = -U_{1,1}$ ) corresponds to the change of the wavenumber due to coupling. We can see that $U_{1,1}$ differs from zero only in the region of the second diaphragm. Dependence of $U_{-1,1}$ ( $U_{1,-1} \approx U_{-1,1}^*$ ) on $z$ is much more complicated.

Results of calculations, presented above, confirm that the one mode representation (10) describes the electromagnetic fields with small errors in the homogeneous part of the section. In the couplers the accuracy is worse, but it is small enough to be used in calculations. For more correct description of fields in the couplers we are need to use more modes as the basis of representation.

3.2 Inhomogeneous structures

We consider the structure, the main parameters of which are given in Table 1. and the dependencies of geometric dimensions are presented in Figure 16 and Figure 17.

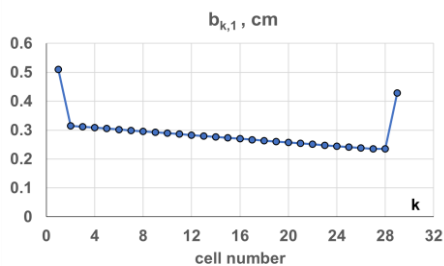
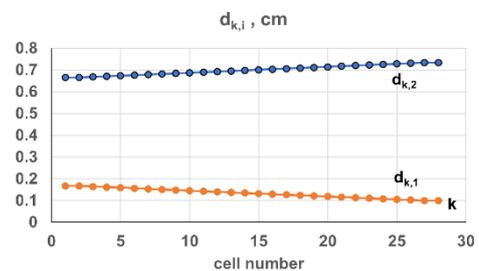

Figure 16 Distributions of disk hole radii

Figure 17 Distributions of disk thicknesses and cell lengths

For non-uniform DLV, there is no procedure for selecting a set of resonator radii that creates a given phase distribution. Therefore, after preliminary selection of the resonator dimensions, it is necessary to carry out a tuning procedure that should give us the required electric field distribution. Today, two tuning methods are most commonly used: the phase Ph-method and the S-method. In the first method, the phase shifts between the resonators are tuned to the required values by a small change in the radii of the cavities (see, for example, [16]). In the second method the combinations of field meanings in some points of several cells are reduced to the desired values by the same actions [17].

For the DLW with phase shift $\varphi = 2\pi/3$ the tuning condition has a form

$$\left|\text{Re}\left(S^{(k)}\right)\right| \Rightarrow \min, \tag{11}$$

where

$$S^{(k)} = -\frac{E^{(k-1)} + E^{(k)} + E^{(k+1)}}{\sqrt{3}\,E^{(k)}}. \tag{12}$$

$E^{(k)}$ is a complex amplitude in the middle of the $k$-th second region $E^{(k)} = E_z\left(r=0, z = z_k + d_{k,1} + d_{k,2}/2\right)$ (see Figure 2).

For tuned homogeneous waveguide $S^{(k)}$ have zero real parts

$$S^{(k)} = i\alpha d, \tag{13}$$

where $\alpha$ is an attenuation coefficient.

For more details of the S method and its restriction, see [18,19].

For tuning our model structure, we used both methods. Results of tuning are represented in Figure 18 and Figure 19. Despite the differences in values of S parameters, the phase distributions are good enough (Figure 19Figure 18). The reflection coefficients for these structures are $|R_{in}| = 8.7\text{E-}3$ (S), and $|R_{in}| = 7.8\text{ E-}3$ (Ph)

Bellow we will consider the section with Ph tuning (see Figure 20).

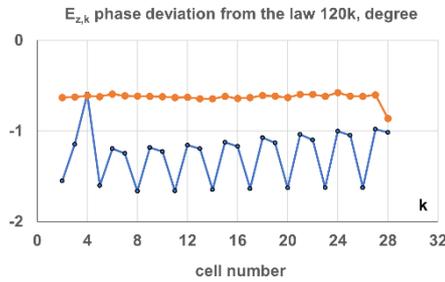

Figure 18 Deviation of phase distributions in the centers of cells from the law 120k; 1-the S tuning method, 2-the Ph tuning method

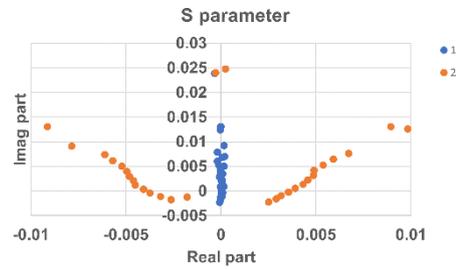

Figure 19 Values of S-parameter; 1-the S tuning method, 2-the Ph tuning method

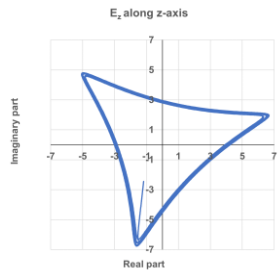

Figure 20 $E_z(r=0,z)$, calculated on the base of the CASCIE code, for the case of tuning by Ph method

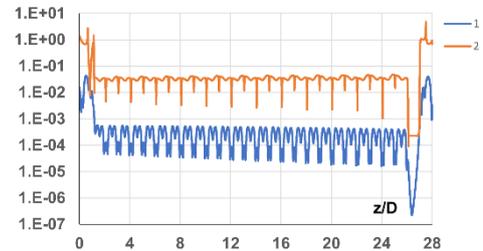

Figure 21 Modulus of relative error in the representation of the longitudinal electric field $E_z(r=0,z)$, calculated using the CASCIE code, by the field of the single mode approach $E_{z,1}(r=0,z)$ (1) and by the field $E_{z,1}^+(r=0,z)$ (2)

Results of calculations, presented in Figure 21, show that the single-mode approximation (10) gives a good accuracy of representation the field distribution in the section under consideration. The relative error between the longitudinal electric field $E_z(r=0,z)$, calculated using the CASCIE code, and the field of the single mode approach $E_{z,1}(r=0,z)$ is less than 1E-3 (for the couplers, see subsection 3.1).

As follows from expression (10), the single mode field is divided into two components: the first, associated with the forward eigen wave and the second, associated with the backward eigen wave. If we limit ourselves to only the first part of the field (forward single mode approximation), then the error increases significantly and for inhomogeneities under consideration can reach 5 percent (see Figure 21).

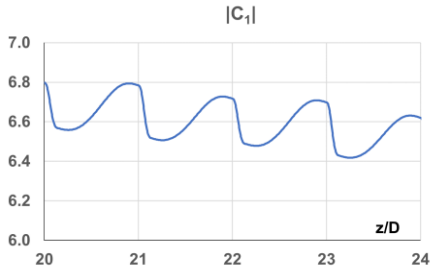

Figure 22 Dependence of modulus of $C_1$ on z-coordinate

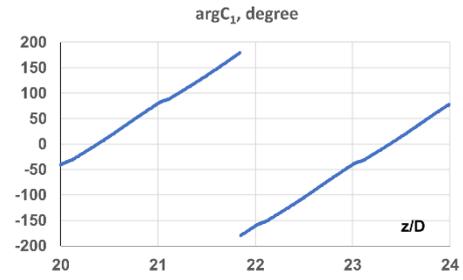

Figure 23 Dependence of argument of $C_1$ on z-coordinate

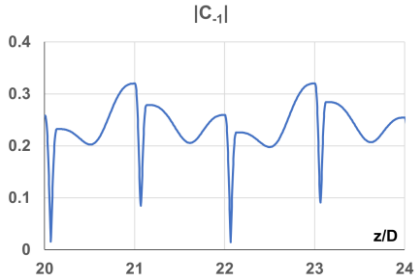

Figure 24 Dependence of modulus of $C_{-1}$ on z-coordinate

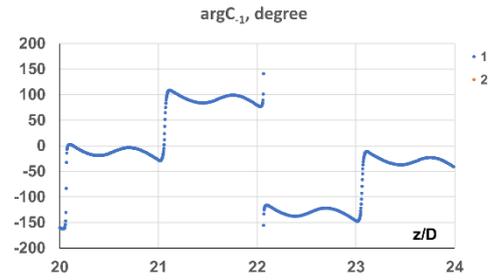

Figure 25 Dependence of argument of $C_{-1}$ on z-coordinate

From these results it follows that the second component in the single-mode approximation can play an important role in representing the field in the accelerating sections. This component is related with the backward eigen wave, but it doesn't represent the backward wave. Indeed, if the "amplitudes" $C_1$ of the first component behaves as an inhomogeneous wave travelling forward (see Figure 22 and Figure 23, phase grows linearly with the z coordinate) then the "amplitudes" $C_{-1}$ doesn't behave as an inhomogeneous wave travelling backward. $C_{-1}$ has a complex phase variation: non-monotonic in the resonator and strong increasing across the disc (see Figure 24 and Figure 25). This behavior is due to the coupling of two "amplitudes" (see Eqs. (5)).

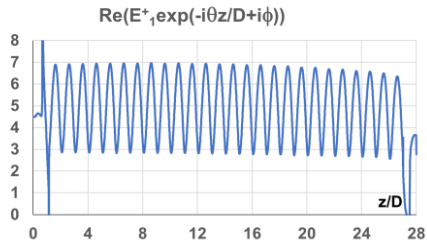

Figure 26 Dependence of longitudinal force $F_1^{(+)}$ on z-coordinate, $\phi = \pi$

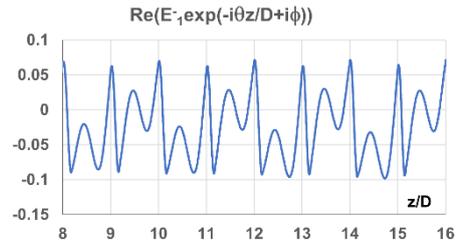

Figure 27 Dependence of longitudinal force $F_1^{(-)}$ on z-coordinate, $\phi = \pi$

The force acting on a relativistic particle as it passes through a section can also be represented as a sum of two components

$$F(z) = F_1^{(+)}(z) + F_1^{(-)}(z) = q\,\text{Re}\left\{E_1^{(+)}\exp\left(-i\frac{\omega z}{c}+i\phi\right)\right\} + q\,\text{Re}\left\{E_1^{(-)}\exp\left(-i\frac{\omega z}{c}+i\phi\right)\right\}, \qquad (14)$$

where $\phi$ is defined by the time of entry into the sections.

Dependences of longitudinal forces $F_1^{(+)}$ and $F_1^{(-)}$ on z-coordinate are presented in Figure 26 and Figure 27. We can see that $F_1^{(+)}$ is a periodic function of $z$ with a period that coincides with the cell length. At the same time, the function $F_1^{(-)}$ has a period equal $2D$.

Evaluation of the integrals of these functions shows that the contribution $F_1^{(-)}$ in energy gain is approximately 1.5%. In many cases this is a significant value.

As $dg_i^{(z)}/dz \neq 0$ along the section then $U_{s',s} \neq 0$. Dependencies of $U_{s',s}$ on $z$ in middle of section are given in Figure 28 and Figure 29. We can see that $U_{1,1}$ differs significantly from zero only in the area of the diaphragms. Coupling coefficients $U_{-1,1}$ also has its maximum values in the area of the diaphragms too, in the area of resonators it has much smaller values.

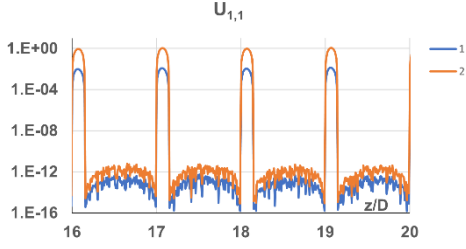

Figure 28 Dependence of $U_{1,1}$ on $z$; 1-real part, 2-imaginary part

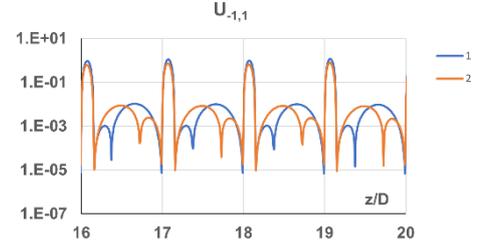

Figure 29 Dependence of $U_{-1,1}$ on $z$; 1-real part, 2- imaginary part

Presented above results confirm that the modified homogeneous eigen waves can be used as a basis for representation of electromagnetic fields in inhomogeneous accelerating sections. Using this basis, we can get a system of coupling equations. Results of study these equations will be given in the next papers.

## 4. CONCLUSIONS

New generalization of the theory of coupled modes, proposed in [1], gives possibility to describe nonuniform accelerating structures. Based on a set of eigen waves of a homogeneous periodic waveguide, a new basis of vector functions is introduced that takes into account the non-periodicity of the waveguide. Representing the total field as the sum of these functions with unknown scalar coefficients, a system of coupled equations that determines the dependence of these coefficients on the longitudinal coordinate was obtained. In some cases we can reduce this system and use the single mode approximation.

In this work we showed that for the studied accelerating section the single wave approximation gives good results. Within the framework of single wave approximation, the fields are represented as sum of two components, one of which is associated with the forward eigen wave, and the second with the backward eigen wave. But this second component is not a backward wave. The distributed coupling of waves due to the presence of distributed reflections leads to the fact that the characteristics of this component are determined by the characteristics of the first component, which is much larger than the second.

The calculation results showed that the coupling coefficients in the equations of coupled modes theory have a complex dependency on $z$. Therefore, solving the coupling equations will require precise numerical methods.

In used approach we represent the original smooth field, satisfying certain boundary conditions, as a sum of more complex vector functions. This could be considered as a drawback of the proposed approach. But this basis is orthonormal and can be expected to be complete Moreover, each elementary vector functions satisfy the same boundary conditions as the original field. These are exactly properties that made it possible to get the system of coupling equations and reduce it. We expect that the two generalized modes (four components in the sums) will be enough to describe the field distributions in the accelerator sections that are used today in accelerators. In this work we presented results of using only one generalized mode (two components in the sums). Adding additional components requires the development of numerical methods for accurately calculating the characteristic of high evanescent waves. One of the positive features of proposed coupled modes theory is the simple and clear procedure for taking into account the beam loading and the simple transition to the case of a homogeneous waveguide.

There are several models, based on the exact electrodynamic approach, for considering beam loading in inhomogeneous structures [4,6,7,8]. The main drawback of these models is their application only to non-periodic structured waveguides based on cylindrical elements and the impossibility of transition to the case of a homogeneous waveguide. In this work we used the model with cylindrical elements too. This is due to the presence of a CASCIE code that is based on this geometry. The use of any other codes, on the basis of which it is possible to calculate the characteristics of homogeneous periodic waveguides, makes it possible to describe inhomogeneous structures with different segment geometries. The only restriction is that the boundary of the periodic waveguide must be fully determined by the finite set of the geometrical parameters.

It is possible that the need to use more complex basis vector functions than their sum is the price to pay for correctly accounting for inhomogeneity.

## APPENDIX 1

The behavior of electromagnetic field is governed by Maxwell's equations

$$rot\,\vec{E} = i\omega\mu_0\vec{H}, \quad (15)$$

$$rot\,\vec{H} = -i\omega\varepsilon_0\varepsilon\vec{E} + \vec{j}. \quad (16)$$

We will look for a solution of equations (15) and (16) in the form of such a series

$$\vec{H} = \sum_{s=-\infty}^{\infty} C_s \vec{H}_s^{(e,z)}. \quad (17)$$

Using the vector relation

$$rot\left(C_s \vec{H}_s^{(e,z)}\right) = \frac{dC_s}{dz}\left[\vec{e}_z \vec{H}_s^{(e,z)}\right] + C_s\left(-i\omega\varepsilon\varepsilon_0 \vec{E}_s^{(e,z)} - \gamma_s^{(z)}\left[\vec{e}_z \vec{H}_s^{(e,z)}\right] + \vec{H}_s^{(\nabla)}\right) \quad (18)$$

and assuming that the series (17) can be differentiated term by term, we get from (16)

$$\vec{E} = -\frac{1}{i\omega\varepsilon_0\varepsilon}rot\,\vec{H} + \frac{\vec{j}}{i\omega\varepsilon_0\varepsilon} = \sum_s C_s \vec{E}_s^{(e,z)} + \frac{\vec{j}}{i\omega\varepsilon_0\varepsilon} - \frac{1}{i\omega\varepsilon_0\varepsilon}\sum_{s>0}\left\{\left(\frac{dC_s}{dz} - \gamma_s^{(z)}C_s\right)\left[\vec{e}_z \vec{H}_s^{(e,z)}\right] + \vec{H}_s^{(\nabla)}C_s\right\}. \quad (19)$$

Suppose that

$$\sum_{s>0}\left\{\left(\frac{dC_s}{dz} - \gamma_s^{(z)}C_s\right)\left[\vec{e}_z \vec{H}_s^{(e,z)}\right] + \vec{H}_s^{(\nabla)}C_s\right\} = \vec{j}_\perp, \quad (20)$$

where $\vec{j} = \vec{j}_l + \vec{j}_\perp$. Then (19) takes the form

$$\vec{E} = \sum_s C_s \vec{E}_s^{(e,z)} + \frac{\vec{j}_l}{i\omega\varepsilon_0\varepsilon} \quad (21)$$

We cannot use in (20) the full current $\vec{j}$ as the sums don't have the longitudinal components.

Substitution of (21) and (17) into the equation (15) gives

$$\sum_{s>0}\left\{\left(\frac{dC_s}{dz} - \gamma_s^{(z)}C_s\right)\left[\vec{e}_z \vec{E}_s^{(e,z)}\right] + \vec{E}_s^{(\nabla)}C_s\right\} = -\frac{1}{i\omega\varepsilon_0\varepsilon}rot\,\vec{j}_l \quad (22)$$

Making some transformations with Eq.(20) and (22) [1,3], we obtain

$$\frac{dC_s}{dz} - \gamma_s^{(e,z)}C_s + \frac{1}{2N_s^{(e,z)}}\frac{dN_s^{(e,z)}}{dz}C_s + \sum_{s'=-\infty}^{\infty} C_{s'}U_{s',s}^{(z)} = \frac{1}{N_s^{(e,z)}}\int_{S_\perp^{(z)}} \vec{j}\vec{E}_{-s}^{(e,z)}dS, \quad (23)$$

where

$$U_{k',k}^{(z)} = \frac{1}{2N_s^{(e,z)}}\sum_i \frac{dg_i^{(z)}}{dz}\int_{S_\perp^{(z)}(z)}\left\{\left[\frac{\partial \vec{E}_{k'}^{(e,z)}}{\partial g_i^{(z)}}\vec{H}_k^{(e,z)}\right] + \left[\frac{\partial \vec{E}_k^{(e,z)}}{\partial g_i^{(z)}}\vec{H}_{k'}^{(e,z)}\right] - \left[\vec{E}_k^{(e,z)}\frac{\partial \vec{H}_{k'}^{(e,z)}}{\partial g_i^{(z)}}\right] - \left[\vec{E}_{k'}^{(e,z)}\frac{\partial \vec{H}_k^{(e,z)}}{\partial g_i^{(z)}}\right]\right\}\vec{e}_z dS, \quad (24)$$

## APPENDIX 2 EIGEN WAVES

Electromagnetic fields in a periodic structured waveguide are the solutions of the Maxwell's equations

$$rot\,\vec{E} = i\omega\mu_0\vec{H}, \quad (25)$$

$$rot\,\vec{H} = -i\omega\varepsilon_0\varepsilon\vec{E}, \quad (26)$$

with the periodic boundary conditions on the side metallic surface of the waveguide

This system has a set of solutions (eigen waves) $\left(\vec{E}_s, \vec{H}_s\right) = \left(\vec{E}_s^{(e)}, \vec{H}_s^{(e)}\right)\exp(\gamma_s z)$, $\gamma_{-s} = -\gamma_s$. Eigen vectors $\left(\vec{E}_s^{(e)}, \vec{H}_s^{(e)}\right)$ are the solutions of such equations

$$\begin{aligned}rot\,\vec{E}_s^{(e)} + \gamma_s\left[\vec{e}_z \vec{E}_s^{(e)}\right] &= i\omega\mu_0\vec{H}_s^{(e)}\\ rot\,\vec{H}_s^{(e)} + \gamma_s\left[\vec{e}_z \vec{H}_s^{(e)}\right] &= -i\omega\varepsilon\varepsilon_0 \vec{E}_s^{(e)}\end{aligned}, \quad (27)$$

with the periodic boundary conditions

$$\vec{E}_{s,\tau}^{(e)} = 0, \quad \vec{H}_{s,\perp}^{(e)} = 0, \quad (28)$$

and satisfy the orthogonality condition

$$\int_{S_t(z)}\left\{\left[\vec{E}_s^{(e)}\vec{H}_{-s'}^{(e)}\right] - \left[\vec{E}_{-s'}^{(e)}\vec{H}_s^{(e)}\right]\right\}\vec{e}_z dS = \begin{cases}0, & s \neq s',\\ sign(s)N_s, & s = s',\end{cases} \quad (29)$$

where $S_\perp(z)$ - any cross-section of the waveguide. $S_\perp(z)$ and $\left(\vec{E}_s^{(e)}, \vec{H}_s^{(e)}\right)$ are the periodic functions of the $z$ coordinate.

Suppose that the boundary of the periodic waveguide is fully determined by the finite set of the geometrical parameters $g_i$, $i=1,...,I$. Eigen solutions $\left(\vec{E}_s^{(e)}, \vec{H}_s^{(e)}\right)$ and propagation constant $\gamma_s$ depend on the geometry of the boundary, therefore we can write $\vec{E}_s^{(e)} = \vec{E}_s^{(e)}\left(\vec{r}_\perp, z, g_1,...,g_I\right)$, $\vec{H}_s^{(e)} = \vec{H}_s^{(e)}\left(\vec{r}_\perp, z, g_1,...,g_I\right)$, $\gamma_{\pm s}(g_1,...,g_I)$.

For each fixed $z$ there are a set of $g_i^{(l)}$ (we will call them local geometrical parameters), that determine the geometry of the cross section $S_\perp^{(z)}(z, g_i^{(l)})$. The remaining parameters we will call global $g_i^{(g)}$. The division of the geometrical parameters $g_i$ into local and global ones depends on $z$.

We can consider new vector functions $\vec{E}_s^{(e,z)} = \vec{E}_s^{(e)}\left(\vec{r}_\perp, z, g_1^{(z)}(z),...,g_I^{(z)}(z)\right)$, $\vec{H}_s^{(e,z)} = \vec{H}_s^{(e)}\left(\vec{r}_\perp, z, g_1^{(z)}(z),...,g_I^{(z)}(z)\right)$, where $g_i^{(z)}(z)$ and its derivatives are continuous functions of $z$. We also introduce a new function $\gamma_s^{(z)} = \gamma_s\left(g_1^{(z)}(z),...,g_I^{(z)}(z)\right)$. The set $g_i^{(z)}(z)$ doesn't describe any real waveguide. But for each fixed $z$ the vectors $\vec{E}_{\pm s}^{(e,z)}$, $\vec{H}_{\pm s}^{(e,z)}$ represent the fields of a periodic waveguide in the cross section $S_\perp^{(z)}(z, g_i^{(l,z)}(z))$, where $g_i^{(l,z)}(z)$ are the set of local geometrical parameters.

The vector functions $\vec{E}_{\pm s}^{(e,z)}$, $\vec{H}_{\pm s}^{(e,z)}$ are no longer the solutions to Maxwell equations. Indeed, as

$$\frac{\partial \vec{E}_s^{(e,z)}}{\partial z} = \frac{\partial \vec{E}_{s,g}^{(e,z)}}{\partial z} + \sum_i \frac{\partial \vec{E}_s^{(e,z)}}{\partial g_i} \frac{dg_i}{dz}, \tag{30}$$

where $\vec{E}_{s,g}^{(e,z)} = \vec{E}_s^{(e,z)}\left(\vec{r}, g_i = const\right)$, then

$$rot\,\vec{E}_s^{(e,z)} = i\omega\mu_0 \vec{H}_s^{(e,z)} - \gamma_s^{(z)}\left[\vec{e}_z \vec{E}_s^{(e,z)}\right] + \vec{E}_s^{(\nabla)}, \tag{31}$$

$$rot\,\vec{H}_s^{(e,z)} = -i\omega\varepsilon\varepsilon_0 \vec{E}_s^{(e,z)} - \gamma_s^{(z)}\left[\vec{e}_z \vec{H}_s^{(e,z)}\right] + \vec{H}_s^{(\nabla)}, \tag{32}$$

where

$$\begin{aligned}\vec{E}_s^{(\nabla)} &= \sum_i \left[\vec{e}_z \frac{\partial \vec{E}_s^{(e,z)}}{\partial g_i}\right] \frac{dg_i}{dz}, \\ \vec{H}_s^{(\nabla)} &= \sum_i \left[\vec{e}_z \frac{\partial \vec{H}_s^{(e,z)}}{\partial g_i}\right] \frac{dg_i}{dz}, \end{aligned} \tag{33}$$

New vector functions $\vec{E}_s^{(e,z)}, \vec{H}_s^{(e,z)}$ still satisfy the boundary conditions on the contour $\vec{r}_\perp = \vec{r}_\perp^{(z)}(z, g_i^{(l,z)}(z))$, which limits the cross section $S_\perp^{(z)}(z, g_i^{(l,z)}(z))$ of some waveguide, and the orthogonality conditions

$$\int_{S_\perp^{(z)}(z)} \left\{\left[\vec{E}_s^{(e,z)} \vec{H}_{-s'}^{(e,z)}\right] - \left[\vec{E}_{-s'}^{(e,z)} \vec{H}_s^{(e,z)}\right]\right\} \vec{e}_z dS = \begin{cases} 0, & s \neq s', \\ sign(s) N_s^{(e,z)}\left(g_i^{(z)}(z)\right), & s = s'. \end{cases} \tag{34}$$

For non-periodic circular disk-loaded waveguide with rectangular geometry we will introduce the notion of the elementary cell with a number $k$ which starts at $z_k = \sum_i^{k-1}(d_{i,1} + d_{i,2})$ and consists of a disk ($z_k < z < z_k + d_{k,1}$, thickness $g_{k,1} = d_{k,1}$, the radius of opening $g_{k,2} = b_{k,1}$) and the following segment of a circular waveguide ($z_k + d_{k,1} < z < z_k + d_{k,1} + d_{k,2}$, the radius $g_{k,3} = b_{k,2}$, length $g_{k,4} = d_{k,2}$) (see Figure 2). For this waveguide at $z_k < z < z_k + d_{k,1}$ there are two local parameters $g_{k,1}^{(l)} = b_{k,1}, g_{k,2}^{(l)} = d_{k,1}$; at $z_k + d_{k,1} < z < z_k + d_{k,1} + d_{k,2}$ the local parameters are $g_{k,3}^{(l)} = b_{k,2}, g_{k,4}^{(l)} = d_{k,2}$. The vector functions $\vec{E}_s^{(e,z)}$, $\vec{H}_s^{(e,z)}$ will fulfill the orthogonality conditions and the boundary conditions on the surface of the considered waveguide when the value of corresponding function $g_i^{(l,z)}(z)$ equals the value of the local parameter $g_{k,i}^{(l)}$. The values of global parameters can be chosen arbitrary.

For circular disk-loaded waveguide with rectangular geometry at $z_k < z < z_k + d_{k,1}$ the local parameters $g_{k,1}^{(l)} = b_{k,1}$, $g_{k,2}^{(l)} = d_{k,1}$ are constant, hence the functions $g_1^{(z)}(z), g_2^{(z)}(z)$ should be constant on this interval too: $g_1^{(z)} = b_{k,1}, g_2^{(z)} = d_{k,1}$. The functions $g_3^{(z)}(z), g_4^{(z)}(z)$ can be arbitrary, but there must be continues transition (including a derivative) from $g_3^{(z)}(z_k - 0), g_4^{(z)}(z_k - 0)$ to $g_3^{(z)}(z_k + d_{k,1} + 0), g_4^{(z)}(z_k + d_{k,1} + 0)$.

At $z_k + d_{k,1} < z < z_k + d_{k,1} + d_{k,2}$ the same is applied to the functions $g_1^{(z)}(z), g_2^{(z)}(z)$ and the local parameters $g_{k,3}^{(l)} = b_{k,2}, g_{k,4}^{(l)} = d_{k,2}$

Therefore, we need to connect two horizontal lines with a function that has as many zero derivatives as possible at both ends of the segment. If we want to have $N$ zero derivatives, we can use a polynomial of degree ($2N + 1$).

We studied the possibility of using several transitions functions. With using polynomial functions, we can represent $g_3^{(z)}(z)$ as:

the polynomial of degree 5

$$g_3^{(z)}(z) = \begin{cases} b_{k-1,2} - (b_{k-1,2} - b_{k,2})\left\{10\dfrac{\tilde{z}^3}{d_{k,1}^3} - 15\dfrac{\tilde{z}^4}{d_{k,1}^4} + 6\dfrac{\tilde{z}^5}{d_{k,1}^5}\right\}, \tilde{z} = z - z_{k,1}, \; z_{k,1} < z < z_{k,2}, \\ b_{k,2}, \hspace{6cm} z_{k,2} < z < z_{k+1,1}, \end{cases} \quad (35)$$

the polynomial of degree 3

$$g_3^{(z)}(z) = \begin{cases} b_{k-1,2} - (b_{k-1,2} - b_{k,2})\left\{3\dfrac{\tilde{z}^2}{d_{k,1}^2} - 2\dfrac{\tilde{z}^3}{d_{k,1}^3}\right\}, \tilde{z} = z - z_{k,1}, \; z_{k,1} < z < z_{k,2}, \\ b_{k,2}, \hspace{6cm} z_{k,2} < z < z_{k+1,1}, \end{cases} \quad (36)$$

We also used a non-polynomial transition

$$g_3^{(z)}(z) = \begin{cases} b_{k-1,2} - (b_{k-1,2} - b_{k,2})\sin\left\{\dfrac{\pi}{2}\left(\dfrac{\tilde{z}}{d_{k,1}}\right)^2\right\}, \tilde{z} = z - z_{k,1}, \; z_{k,1} < z < z_{k,2}, \\ b_{k,2}, \hspace{6cm} z_{k,2} < z < z_{k+1,1}, \end{cases} \quad (37)$$

Results are shown in Figure 30 and Figure 31. Dependences of $g_3^{(z)}(z)$ on the z-coordinate in the vicinity of the second diaphragm ( $1 < z/D < 1.2$ ), where the gradients take its maximum value, for different transition functions are given in Figure 30.

From Figure 31, where modulus of relative error are given for the initial part of model section (see subsection 3.2), we can conclude that the form of the transition function weakly effects on the calculation results.

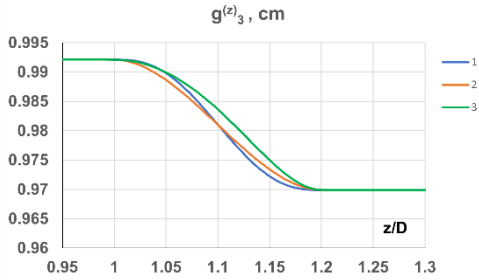
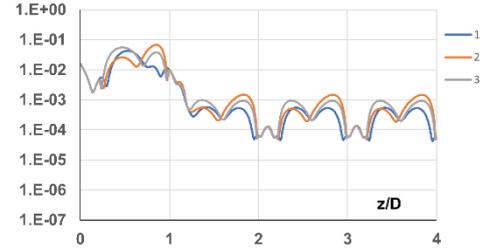

*Figure 30 Dependence of $g_3^{(z)}(z)$ on the z-coordinate in the vicinity of the second diaphragm ( $1 < z/D < 1.2$ ) for different transition functions: 1- the polynomial of degree 5;2- the polynomial of degree 3, 3-the non-polynomial transition*

*Figure 31 Modulus of relative error in the representation of the longitudinal electric field $E_z(r=0,z)$ , calculated on the base of the CASCIE code, by the field of the single mode approach $E_{z,1}(r=0,z)$ for different transition functions: 1- the polynomial of degree 5;2- the polynomial of degree 3, 3-the non-polynomial transition*

# REFERENCES

1 M.I. Ayzatsky, Coupled-mode theory for non-periodic structured waveguides, https://doi.org/10.48550/arXiv.2309.06280, 2024.
2 LA.Weinstein, Electromagnetic waves, Moscow: Radio i svayz; 1988, (In Russian).
3 M.I. Ayzatsky, Representation of fields in inhomogeneous structured waveguides, https://doi.org/10.48550/arXiv.2404.00708, 2024.
4 S. A. Heifets, S. A. Heifets, Longitudinal electromagnetic fields in an aperiodic structure, IEEE Trans.Microwave Theor.Tech., 1994, V.42 , N.1, pp.108 – 117; SLAC-PUB-5907,1992
5 U.V.Rienen, Higher order mode analysis of tapered disc-loaded waveguides using the mode matching technique, Particle Accelerators, 1993, Vol. 41, pp. 173-201
6 S. Amari, J. Bornemann, and R. Vahldieck, Accurate analysis of scattering from multiple waveguide discontinuities using the coupled integral equation technique, J. Electromag. Waves Applicat., 1996, V.10, pp. 1623–1644.
7 V.A.Dolgashev, Calculation of impedance for multiple waveguide junction using scattering matrix formulation, Proceedings of the International Computational Accelerator Physics Conference (ICAP)98, Monterey,1998.
8 M.I. Ayzatsky, Modification of coupled integral equations method for calculation the accelerating structure characteristics, PAST, 2022, N.3, pp.56-61, https://doi.org/10.46813/2022-139-056; https://doi.org/10.48550/arXiv.2203.035182022.
9 M.I. Ayzatsky, Fast code CASCIE (Code for Accelerating Structures -Coupled Integral Equations). Test Results, https://doi.org/10.48550/arXiv.2209.11291, 2022.
10 A. Grudiev, W. Wuensch, Design of the CLIC main linac accelerating structure for CLIC conceptual design report, Proceedings of LINAC2010, pp.211-213.
11 Christopher Nantista, Sami Tantawi, and Valery Dolgashev, Low-field accelerator structure couplers and design technique, PhysRevSTAB, V7, 072001 (2004), https://doi.org/10.1103/PhysRevSTAB.7.072001
12 C. Serpico, A. Grudiev , R. Vescovo, Analysis and comparison between electric and magnetic power couplers for accelerators in Free Electron Lasers (FEL), Nuclear Instruments and Methods in Physics Research, A823, 2016, pp.8-14, http://dx.doi.org/10.1016/j.nima.2016.06.131.
13 D.H. Whittum, Introduction to Electrodynamics for Microwave Linear Accelerators, In: S.I.Kurokawa, M.Month, S.Turner (Eds) Frontiers of Accelerator Technology, World Scientific Publishing Co.Pte.Ltd., 1999.
14 M.I. Ayzatsky A novel approach to the synthesis of the electromagnetic field distribution in a chain of coupled resonators, PAST, 2018, N.3, pp.29-37.
15 David Alesinia, Alessandro Galloa, Bruno Spataroa, Agostino Marinellib, Luigi Palumbo, Design of couplers for traveling wave RF structures using 3D electromagnetic codes in the frequency domain, Nuclear Instruments and Methods in Physics Research, 2007, A 580, pp.1176–1183.
16 R.B Neal, General Editor, The Stanford Two-Mile Accelerator, New York, W.A. Benjamin, 1968
17 T.Khabiboulline, V.Puntus, M.Dohlus et al, A new tuning method for traveling wave structures. Proceedings of PAC95, pp.1666-1668; T.Khabiboulline, M.Dohlus, N.Holtkamp, Tuning of a 50-cell costant gradient S-band travelling wave accelerating structure by using a nonresonant perturbation method, Internal Report DESY M-95-02, 1995.
18 M.I. Ayzatskiy, V.V. Mytrochenko, Electromagnetic fields in nonuniform disk-loaded waveguides, PAST, 2016, N.3, pp.3-10; M.I. Ayzatsky, V.V. Mytrochenko, Electromagnetic fields in nonuniform disk-loaded waveguides, https://arxiv.org/abs/1503.05006, 2015.
19 M.I.Ayzatsky, V.V.Mytrochenko, Numerical design of nonuniform disk-loaded waveguides, https://arxiv.org/abs/1604.05511, 2016; M.I.Ayzatsky, V.V.Mytrochenko, Numerical investigation of tuning method for nonuniform disk-loaded waveguides, https://arxiv.org/abs/ 1606.04292, 2016.